\documentclass[preprint,showpacs,preprintnumbers,amsmath,amssymb]{revtex4} 
\usepackage{graphicx}
\usepackage{dcolumn}
\usepackage{bm}%

\begin{document}

\title{\bf A General Relativistic study of the neutrino path and calculation
of minimum photosphere for different stars}

\author{Ritam Mallick$^{a,b}$\footnote{Email : ritam@bosemain.boseinst.ac.in}  
 and Sarbani Majumder$^{a,b}$}

\affiliation{$^a$Department of Physics; Bose Institute;
 93/1, A.P.C Road; Kolkata - 700009; INDIA}
\affiliation{$^b$Centre for Astroparticle Physics and
Space Science; Bose Institute; 93/1, A.P.C Road; Kolkata - 700009;
INDIA}

\vskip 0.4in
\begin{abstract} 
A detailed general relativistic (GR) calculation of the neutrino path 
for a general metric describing a rotating star is studied. We have
calculated the neutrino path along a plane, with the consideration 
that the neutrino does not at any time leave the plane. The expression 
for the minimum photosphere radius (MPR) is obtained and matched with the
Schwarzschild limit. The MPR is calculated for the 
stars with two different equations of state (EOS) each rotating with two 
different velocities. The results shows that the MPR for the hadronic 
star is much greater than
the quark star and the MPR increases as the rotational velocity of the star 
decreases. The MPR along the polar plane is larger than that along the 
equatorial plane.
\end{abstract}

\maketitle

\section{Introduction }

Gamma Ray Bursts (GRB), the possible engines for GRBs and its connection
with the neutrino production is a field of high current interest. 
It was proposed that the neutrino-antineutrino annihilation to 
electron-positron pairs in compact stars is a possible and important 
candidate to explain the energy source of GRBs.
The previous calculations of the reaction $\nu {\bar \nu} 
\rightarrow e^+ e^-$ in the vicinity of a neutron star have 
been based on Newtonian gravity  \cite{key-2,key-3}, {\it i.e.}  
$( 2GM/c^2R ) << 1 $. The effect of gravity was incorporated in 
refs. \cite{key-4,key-5}, but only for a static star.

Neutron stars are objects formed 
in the aftermath of supernova. The central density of these stars can be as 
high as $10$ times that of normal nuclear matter. At such high density, 
any small perturbation,
{\it e.g.} spin down of the star, may trigger the phase transition from 
nuclear to quark matter
system. As a result, the neutron star may fully convert to a quark star or
a hybrid star with a quark core. It has been shown \cite{key-6} that such a 
phase transition \cite{key-7} produces a large amount of high energy neutrinos. 
These neutrinos (and antineutrinos) could annihilate and 
give rise to electron-positron pairs through the reaction $\nu {\bar \nu} 
\rightarrow e^+ e^-$. These $e^+ e^-$ pairs may further give rise to gamma 
rays which may provide a possible explanation of the observed GRB. 
Furthermore, the rotating neutron star
has been shown \cite{key-8} to produce the observed beaming effect. At 
present, it is necessary to have
a better understanding of the energy deposition in the neutrino 
annihilation to $e^{+} e^{-}$ 
in the realistic neutron star environment.

We would like to study the $\nu + \overline{\nu} \rightarrow e^+ + e^- $ 
energy deposition rate near a rotating compact star. This reaction is important 
for the study of gamma ray bursts. The General Relativistic (GR) effect 
increase the efficiency of the process immensely,
but the inclusion of the rotational effect is yet to be incorporated 
and studied. 
The geodesic of neutrinos are also important in the study
of pulse shapes and accretion disc illumination \cite{key-11}
to name a few. Therefore the
path of neutrino (or generally of massless particle) is of immense 
importance and needs a detailed study. The neutrino path for Schwarzschild
metric along the equatorial plane can be found in text books \cite{key-12} 
and different papers \cite{key-4}. Asano and Fukuyama \cite{key-12a,key-12b} 
did the 
same calculation near a thin accretion disc using Kerr metric. Prasanna
and Shrubabati \cite{key-13} studied it for a slowly rotating star.
To address all the above problems we present a detailed GR study of the 
neutrino path for a most general metric describing a rotating star 
\cite{key-14}
along a plane. We have made our calculation using two different EOS, one 
quark and the other hadronic.

In this paper first we will discuss about the metric, the EOS and the star 
structure. Next we will present the detailed GR calculation of the neutrino
path and minimum photosphere. Finally we will present our results for the
two EOS and have a brief discussion.

\section{The star}

The structure of the star is described by Cook-Shapiro-Teukolsky (CST) 
metric \cite{key-14}
\begin{eqnarray}
ds^2 = -e^{\gamma+\rho}dt^2 + e^{2\alpha}(dr^2+r^2d\theta^2) 
+ e^{\gamma-\rho}r^2 sin^2\theta(d\phi-\omega dt)^2.
\end{eqnarray}
Accurate models of rotating neutron stars for tabulated EOS can be computed
numerically using the {\bf 'rns'} code \cite{key-15,key-16,key-17}. 
This computer code
computes the metric functions $\alpha, \gamma, \rho$ and 
$\omega$ appearing in the axisymmetric metric, and these metric functions 
depends only on the coordinates $\theta$ and $r$. The metric function
$\omega$ is the term responsible for the frame dragging effect and would
vanish if the rotational velocity ($\Omega$) is zero. The coordinate $r$ is
related to the standard radial coordinate that appears in the Schwarzschild
metric,$r_s$, by $r_s = r e^{(\gamma - \rho )/2}$ \cite{key-17a}. In the
limit of zero rotation, the following combination of metric functions are
\begin{eqnarray}
Lim_{\omega \rightarrow 0} re^{-\rho}=\frac{r_s}
{\sqrt{1-\frac{2M}{r_s}}} \\
Lim_{\omega \rightarrow 0} e^{(\gamma + \rho)/2}= {\sqrt{1-\frac{2M}{r_s}}} \\
Lim_{\omega \rightarrow 0} e^{\alpha - [(\gamma + \rho)/2]} dr = 
\frac{dr_s}{{1-\frac{2M}{r_s}}}.
\end{eqnarray}
 
We have previously mentioned that tabulated EOS are needed to compute the 
code numerically. In this paper we have computed for two different EOS, the
quark EOS and the hadronic EOS.
The hadronic EOS has been evaluated using the nonlinear Walecka
model \cite{key-18}. The Lagrangian density in this model is given
by: 

\begin{eqnarray}
{\cal L}&=&\sum_{i}\bar{\psi_{i}}(i\gamma^{\mu}\partial_{\mu}-m_{i}
+g_{\sigma i}\sigma+g_{\omega i}\omega_{\mu}\gamma^{\mu}
-g_{\rho i}\rho_{\mu}^{a}\gamma^{\mu}T_{a})\psi_{i}\nonumber \\
&-& \frac{1}{4}\omega^{\mu\nu}\omega_{\mu\nu}
+\frac{1}{2}m_{\omega}^{2}\omega_{\mu}\omega^{\mu}
+\frac{1}{2}(\partial_{\mu}\sigma\partial^{\mu}\sigma
-m_{\sigma}^{2}\sigma^{2})
\nonumber \\
&-&\frac{1}{4}\rho_{\mu\nu}^{a}\rho_{a}^{\mu\nu}+\frac{1}{2}m_{\rho}^{2}\rho_{\mu}^{a}\rho_{a}^{\mu}
-\frac{1}{3}bm_{N}(g_{\sigma N} {\sigma})^{3}-\frac{1}{4}c(g_{\sigma N} 
{\sigma})^{4} \nonumber \\
&+&\bar{\psi_{e}}(i\gamma^{\mu}\partial_{\mu}-m_{e})\psi_{e}
\label{1}
\end{eqnarray}

The Lagrangian in eqn. 5 includes nucleons (neutrons and protons),
electrons, isoscalar scalar, isoscalar vector and isovector vector
mesons denoted by $\psi_{i}$, $\psi_{e}$, $\sigma$, $\omega^{\mu}$
and $\rho^{a,\mu}$, respectively. The Lagrangian also includes cubic
and quartic self interaction terms of the $\sigma$ field. The parameters
of the nonlinear Walecka model are meson-baryon coupling constants,
meson masses and the coefficient of the cubic and quartic self interaction
of the $\sigma$ mesons (b and c, respectively). The meson fields interact
with the baryons through linear coupling. The $\omega$ and $\rho$
meson masses have been chosen to be their physical masses. The rest
of the parameters, namely, nucleon-meson coupling constants 
($\frac{g_{\sigma}}{m_{\sigma}},\frac{g_{\omega}}{m_{\omega}}$
and $\frac{g_{\rho}}{m_{\rho}}$) and the coefficients 
of cubic and quartic terms of the $\sigma$ meson self interaction
(b and c, respectively) are determined by fitting the nuclear matter
saturation properties, namely, the binding energy/nucleon (-16 MeV), 
baryon density
($\rho_{0}$=0.17 $fm^{-3}$), symmetry energy coefficient (32.5 MeV),
Landau mass (0.83 $m_{n}$) and nuclear matter incompressibility (300 MeV).
We have used a stable three-flavour quark matter EOS, 
obtained from the standard Bag model with $B^{1/4} = 145 MeV$. 

The shape of a fast rotating neutron star becomes oblate spheroid 
\cite{key-14}. The star gets compressed along the 
z-axis and along x and y-axes, it bulges by equal amounts, the polar radius is 
thus smaller than equatorial radius.

\section{GR calculation}

The metric is independent of $'t'$ and $'\phi'$, the coordinates are cyclic,
hence the corresponding covariant generalized momenta is constant 
\cite{key-12}, {\it i.e }
\begin{eqnarray}
p_t=p_0=const.=-E  \nonumber \\
p_{\phi}=p_3=const=L
\end{eqnarray}

The magnitude of the 4 vector energy momentum is given by \cite{key-12},
\begin{eqnarray}
g_{ij}p^ip^j + \mu^2 = 0,
\end{eqnarray} 
where $\mu$ is the rest mass of the particle. Writing it explicitly we have
\begin{eqnarray}
g_{00}{p^0}^2 + g_{11}{p^1}^2 + g_{22}{p^2}^2 + g_{33}{p^3}^2 + 
g_{30}p^3p^0 + g_{03}p^0p^3 + \mu^2 = 0. 
\end{eqnarray}

These are contravariant momenta, and
to find the contravariant momenta $p^i$, we calculate the inverse matrix
$g^{\mu \nu}$, of the metric given in eqn. 1

\vskip 0.2in

\maketitle
\begin{tabular*}{0.75\textwidth}{@{\extracolsep{\fill}} | l  l  l  l | }
$-e^{-(\gamma + \rho)}$ & $0$ & $0$ & $-\omega e^{-(\gamma + \rho)}$ \\
$0$ & $e^{-2\alpha}$ & $0$ & $0$ \\
$0$ & $0$ & $\frac{1}{r^2} e^{-2\alpha}$ & $0$ \\
$-\omega e^{-(\gamma + \rho)}$ & $0$ & $0$ &
$-(\omega^2e^{-(\gamma+\rho)} -\frac{e^{(\gamma-\rho)}}{r^2 sin^2\theta}$ \\
\end{tabular*}

\vskip 0.2in

Using the original $g_{\mu \nu}$ and the inverse $g^{\mu \nu}$ matrix we 
calculate the contravariant momenta, which are given by 
\begin{eqnarray}
p^0= g^{00}p_0 + g^{03}p_3 = e^{-(\gamma + \rho)}(E-\omega L)=
e^{-(\gamma + \rho)}B(\omega) \nonumber \\
p^3=g^{30}p_0 + g^{33}p_3 = e^{-(\gamma + \rho)}[\omega B(\omega) + 
\frac{e^{2\rho}}{r^2sin^2\theta}L]
\end{eqnarray}

We consider the particle motion is in a particular $\theta=const$ plane,
and orient the coordinate system such that the particle lies in the
equatorial plane for $\theta=\frac{\pi}{2}$.
The particle has at start, and continues to have zero momentum in the
given plane {\it i.e} $p^{\theta}=p^2=0$.

Finally, substituting these values in the above eqn. 8 we get
\begin{eqnarray}
-e^{-(\gamma + \rho)}B^2 + \frac{L^2}{r^2sin^2\theta}e^{\rho -\gamma } +
e^{2\alpha}(\frac{dr}{d\lambda})^2 + \mu^2 = 0.
\end{eqnarray}
if, $\lambda = \frac{\tau}{\mu} {\it i.e}$ propertime per unit rest mass
\begin{eqnarray}
\frac{dr}{d\lambda}=\frac{dr}{d\tau}.\frac{d\tau}{d\lambda} = \mu
\frac{dr}{d\tau} \nonumber \\
\frac{dr}{d\lambda}= \mu (\frac{dr}{d\phi}.\frac{d\phi}{d\tau}) = \mu (\frac
{dr}{d\phi})(\frac{d\phi}{d\lambda})(\frac{d\lambda}{d\tau}) \nonumber \\
\frac{dr}{d\lambda}= (\frac{dr}{d\phi}).p^{\phi}.
\end{eqnarray}
We define $\overline{E}=\frac{E}{\mu}$ and 
$\overline{L}=\frac{L}{\mu}$. 
As the particle is massless (neutrino), therefore we define
\begin{eqnarray}
Lim_{\mu \rightarrow 0} \frac{\overline{L}}{\overline{E}}=b
\end{eqnarray}
where $b$ is the impact parameter. substituting this in eqn. 10, we have
\begin{eqnarray}
e^{2\alpha}(\frac{dr}{d\phi})^2[\omega(1-\omega b) + \frac{be^{2\rho}}
{r^2sin^2\theta}]^2 - e^{(\gamma + \rho)}(1-\omega b)^2 + 
\frac{b^2}{r^2sin^2\theta}e^{\gamma + 3\rho} = 0.
\end{eqnarray}

\vskip 0.2in
The lagrangian of the system we are considering is given by
\begin{eqnarray}
\L = \frac{1}{2} g_{ij}\dot{x}^i\dot{x}^j
\end{eqnarray}
where, $\dot{x}^i = \frac{dx^i}{d\lambda}$, $\lambda$ the affine parameter.
Therefore writing it explicitly, we get
\begin{eqnarray}
\L = (-e^{\gamma + \rho} + e^{\gamma - \rho}\omega^2r^2sin^2\theta)\dot{t}^2
+e^{2\alpha}\dot{r}^2 +e^{\gamma - \rho}r^2sin^2\theta\dot{\phi}^2 -
2e^{\gamma - \rho}\omega r^2sin^2\theta \dot{\phi}\dot{t}.
\end{eqnarray}
Using the above lagrangian we
can write the covariant momentums, and they are given in terms of total
energy and total angular momentum 
\begin{eqnarray}
p_0=p_t=\frac{\partial L}{\partial \dot{t}} = (-e^{\gamma + \rho} + 
e^{\gamma - \rho}\omega^2r^2sin^2\theta)\dot{t} - e^{\gamma - \rho}
\omega r^2sin^2\theta \dot{\phi} = -E \nonumber  \\
p_3=p_{\phi}=\frac{\partial L}{\partial \dot{\phi}}=e^{\gamma - \rho}
r^2sin^2\theta\dot{\phi} - e^{\gamma - \rho}\omega r^2sin^2\theta \dot{t} = L.
\end{eqnarray}
Having written the momenta in terms of total energy and total angular 
momentum it is quite simple to solve for $\dot{t}$ and $\dot{\phi}$. They 
are given by,
\begin{eqnarray}
\dot{t}=\frac{E - \omega L}{e^{\gamma + \rho}} \nonumber  \\
\dot{\phi}= \frac{L}{e^{\gamma - \rho}r^2sin^2\theta} + \frac{\omega 
(E - \omega L)}{e^{\gamma + \rho}}
\end{eqnarray}

The angle $\theta_r$ between the particle trajectory and the tangent 
vector to the orbit can be derived by constructing the local lorentz 
tetrad ${k^{\mu}}_{\nu}$ for our metric

\vskip 0.2in

\maketitle
\begin{tabular*}{0.75\textwidth}{@{\extracolsep{\fill}} | l  l  l  l | }
$e^{(\gamma + \rho)/2}$ & $0$ & $0$ & $0$ \\
$0$ & $e^{\alpha}$ & $0$ & $0$ \\
$0$ & $0$ & $r e^{\alpha}$ & $0$ \\
$-e^{(\gamma - \rho)/2} \omega r sin\theta$ & $0$ & $0$ & 
$e^{(\gamma - \rho)/2}r sin\theta$ \\
\end{tabular*}

\vskip 0.2in

The angle $\theta_r$ is given by
\begin{eqnarray}
tan\theta_r = \frac{V^1}{V^2} = \frac{{k^1}_r V^r}{{k^3}_t V^t + 
{k^3}_{\phi} V^{\phi}}
\end{eqnarray}
where, $V^r = \frac{\dot{r}}{\dot{t}}$ and $V^{\phi} = \frac{\dot{\phi}}
{\dot{t}}$ are the local velocities. Using eqn. 17 $V^{\phi}$ can be written as
\begin{eqnarray}
V^{\phi}= \frac{L}{E - \omega L}. \frac{e^{2\rho}}{r^2sin^2\theta} + \omega
= A(r,\theta) + \omega.
\end{eqnarray}
Therefore the angle $\theta_r$ is 
\begin{eqnarray}
tan\theta_r = \frac{e^{\alpha}}{e^{(\gamma - \rho)/2} r sin\theta}.
\frac{V^{\phi}}{V^{\phi} - \omega} (\frac{dr}{d\phi}) \nonumber \\
     = \frac{e^{\alpha}}{e^{(\gamma - \rho)/2} r sin\theta}.
\frac{A(r,\theta) + \omega}{A(r,\theta)} (\frac{dr}{d\phi}) 
\end{eqnarray}
Squaring the above equation and writing as
\begin{eqnarray}
(\frac{dr}{d\phi})^2 = [\frac{A}{A+ \omega}]^2
\frac{e^{(\gamma - \rho)} r^2 sin^2\theta}{e^{2\alpha}} tan^2 \theta_r.
\end{eqnarray}
we get the final form of $\frac{dr}{d\phi}$.
Substituting this value in eqn. 13, we get 
\begin{eqnarray}
e^{2\alpha}[\frac{A(r,\theta)}{A(r,\theta)+\omega}]^2
\frac{e^{(\gamma - \rho)} r^2sin^2 \theta}{e^{2\alpha}} tan^2 \theta_r
[\omega (1-\omega b)+\frac{e^{2\rho}b} {r^2sin^2\theta}]^2 \nonumber \\
- e^{(\gamma + \rho)}(1-\omega b)^2 + \frac{b^2}{r^2sin^2\theta }
e^{(\gamma + 3\rho)} = 0.
\end{eqnarray}
This equation can be solved using the potentials obtained from
{\bf rns} code to obtain a minimum 
radius $r=R$, the minimum photosphere radius, below which a 
massless particle (neutrino) emitted
tangentially to the stellar surface ($\theta_R =0$) would be 
gravitationally bound.

In the limit in which CST metric reduces to Schwarzschild metric, {\it i.e}
\begin{eqnarray}
Lim_{\omega \rightarrow 0} re^{-\rho}=\frac{r_s}
{\sqrt{1-\frac{2M}{r_s}}} \nonumber
\end{eqnarray}
eqn. 22 (for $\theta=\pi/2$) 
reduces to the equation for $b$ obtained by Salmonson and Wilson\cite{key-4}, 
{\it i.e} 
\begin{eqnarray}
b=\frac{r_s cos\theta_r}{\sqrt{1-\frac{2M}{r_s}}}. \nonumber
\end{eqnarray}

\section{Results}

\begin{figure}
   \centering
\includegraphics{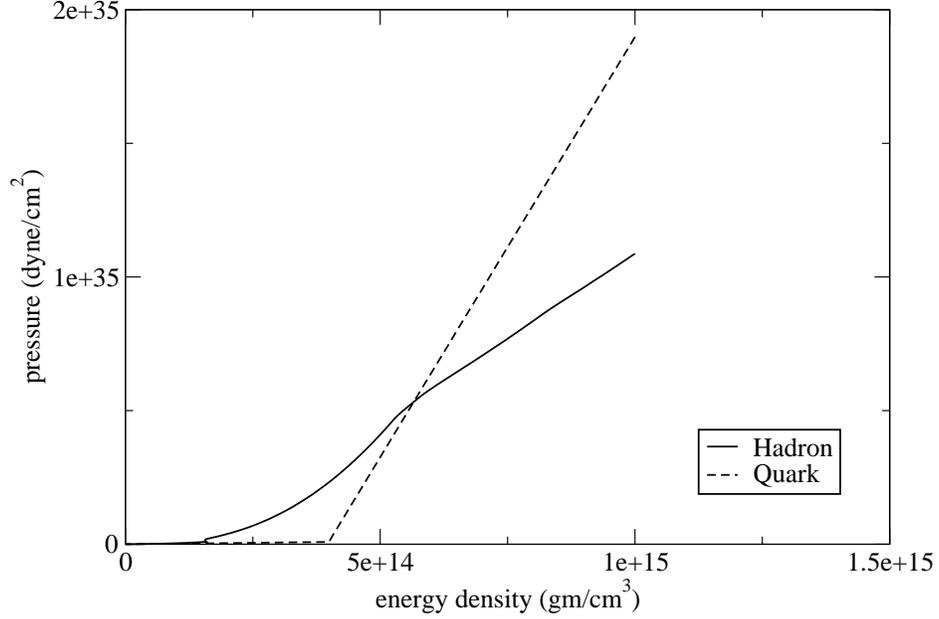}
\caption{Variation of pressure with energy density for quark and hadronic
matter EOS.}
\end{figure}

\begin{figure}
\vskip 0.2in
   \centering
\includegraphics{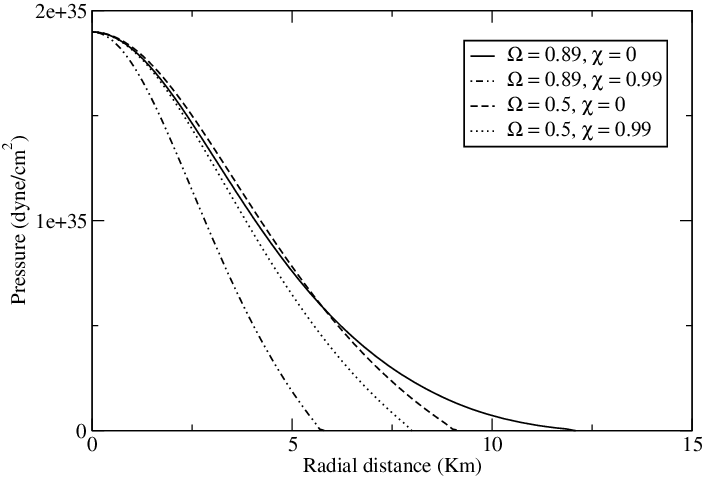}
\caption{Variation of pressure along the radial direction of the star for two 
different rotational velocities each with two different values of $\chi$ for 
the quark matter EOS.}
\end{figure}

\begin{figure}
\vskip 0.2in
   \centering
\includegraphics{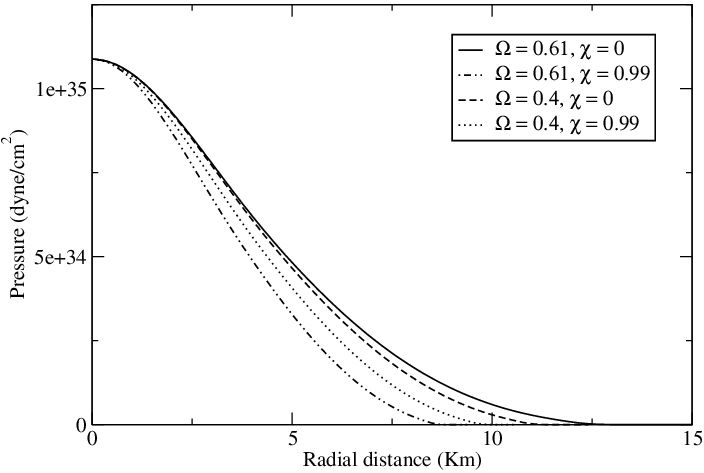}
\caption{Variation of pressure along the radial direction of the star for two
different rotational velocities each with two different values of $\chi$ for
the hadronic matter EOS.}
\end{figure}

The minimum photosphere is calculated solving eqn. 22 using the potential
functions obtained from the { \bf 'rns'} code. Starting our calculation by 
choosing the central energy density of the star to be 
$1 \times 10^{15} gm/cm^3$. Fig. 1, 2, 3, 4 and 5 gives the nature of the 
equations of state used. 
In the figures $\chi=cos\theta$.
Fig. 1 shows that the quark matter EOS, considered here is much
steeper than the hadronic matter EOS. Fig. 2 (for quark matter EOS) shows 
that at the centre of the star the pressure is maximum, and as we go outside 
it falls off, and becomes zero outside the star. Along the pole the pressure 
falls off in a much steeper way than along the equator, as it is of much 
shorter length.
As the the rotational velocity decreases the equatorial radius of the star 
decreases but the polar radius increases (although still less than equatorial 
radius). For the keplerian velocity the star is maximally deformed and as the
rotational velocity of the star decreases the star regains a much spherical 
shape. Fig. 3 show the same nature for a hadronic star. In fig. 4 and 5 the 
variation of energy density is shown for the quark and hadronic star 
simultaneously, and its nature is more or less similar to that of the 
behaviour of pressure discussed above.

\begin{figure}
   \centering
\includegraphics{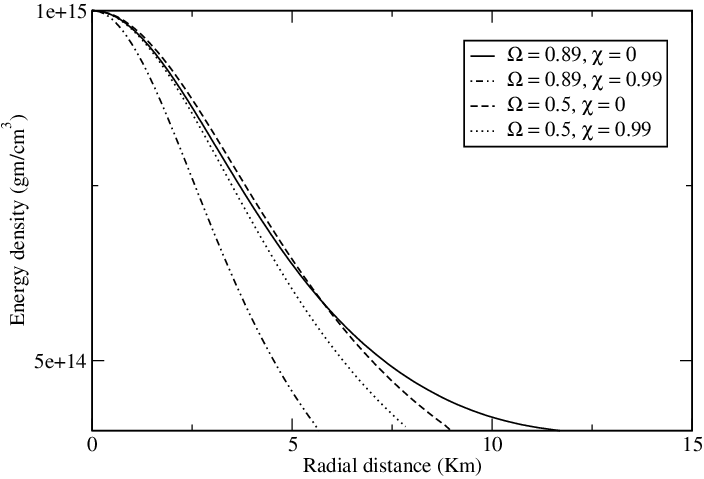}
\caption{Variation of energy density along the radial direction of the 
star for two different rotational velocities each with two different values 
of $\chi$ for the quark matter EOS.}
\end{figure}

\begin{figure}
\vskip 0.2in
   \centering
\includegraphics{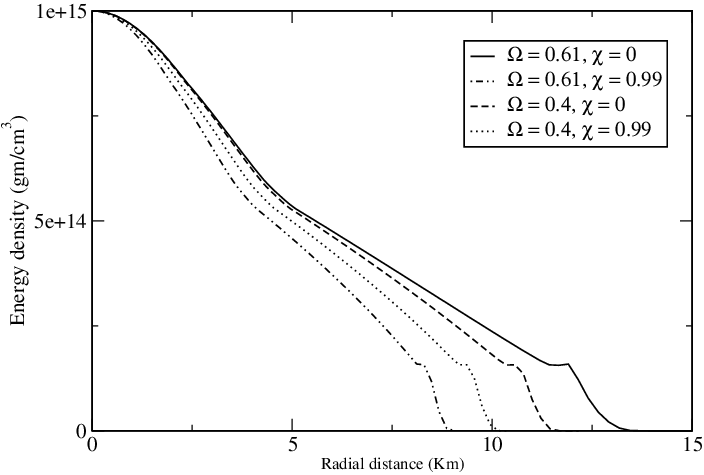}
\caption{Variation of energy density along the radial direction of the star 
for two different rotational velocities each with two different values of 
$\chi$ for the hadronic matter EOS.}
\end{figure}

\vskip 0.3in
\begin{tabular}{|l|l|l|l|l|l|}
\hline
\multicolumn{6}{|c|}{Table 1} \\
\hline
EOS & $\Omega$ & Mass in $M_*$ & $r_e$, $r_p$ in Km 
& $\chi$ & MPR in Km \\ \hline
Quark & $0.89$ & $2.8$ & $12, 5.5$ & $0$ & $2.74$ \\
 & $0.89$ & $2.8$ & $12, 5.5$ & $0.5$ & $2.74$ \\
 & $0.89$ & $2.8$ & $12, 5.5$ & $0.99$ & $3.11$ \\
 & $0.5$ & $2.2$ & $9, 8$ & $0$ & $3.6$ \\
 & $0.5$ & $2.2$ & $9, 8$ & $0.5$ & $3.7$ \\
 & $0.5$ & $2.2$ & $9, 8$ & $0.99$ & $4.96$ \\ \hline
hadron & $0.61$ & $2$ & $16, 9$ & $0$ & $3.9$ \\
 & $0.61$ & $2$ & $16, 9$ & $0.5$ & $4$ \\
 & $0.61$ & $2$ & $16, 9$ & $0.99$ & $5.5$ \\
 & $0.4$ & $1.7$ & $12, 10$ & $0$ & $4.85$ \\
 & $0.4$ & $1.7$ & $12, 10$ & $0.5$ & $4.96$ \\
 & $0.4$ & $1.7$ & $12, 10$ & $0.99$ & $5.85$ \\ \hline
\end{tabular}
\vskip 0.3in

Using the quark matter EOS on the { \bf 'rns'} code, the keplerian velocity
of the quark star comes out to be $0.89 \times 10^4 s^{-1}$. For comparison
we have also computed the { \bf 'rns'} code with rotational velocity
of $0.5 \times 10^4 s^{-1}$. The same treatment is done also for the hadronic
EOS where the keplerian velocity is $0.61 \times 10^4 s^{-1}$ and
for comparison the other rotational velocity was chosen to be
$0.4 \times 10^4 s^{-1}$. The code solves the metric for
the given EOS and gives the different potential functions as a function
of $r$ and $\theta$. Solving eqn. 22 with these values of potential functions
for different $\theta$ we obtain the value of minimum photosphere for
different planes. Table 1. sums up all our results in a compact form.

Let us now analyze the table given above. It points out the fact that
as the rotational velocity decreases the mass of the star also decreases. For
the same central energy density, the mass of the quark star is much greater 
than that of the hadronic star but the radius is much smaller. It 
signifies that the quark matter EOS, considered in our work is much 
steeper than that of hadronic matter EOS as pointed out previously 
in the figures. As the rotational velocity decreases
the equatorial radius decreases but the polar radius increases. It shows that
the star is maximally deformed for the keplerian velocity and as the 
rotational velocity of the star decreases it tries to regain a more 
spherical shape. A static star is of spherical shape, where 
polar and equatorial radius are same.

In the above table we have tabulated the minimum photosphere radius (MPR)
for three different values of $\chi$, { \it i.e} for three planes. Along the
equator ($\chi =0$), along the pole ($\chi=0.99$) and along a plane lying at
$\chi = 0.5$. The MPR is minimum along the equator and maximum along the 
pole. For $\chi=0.5$ it lies somewhere in between these two values. The 
MPR is minimum for the quark star rotating with keplerian velocity and 
is maximum for the hadronic star rotating with $0.4 \times 10^4 s^{-1}$
velocity. The MPR is much greater for the hadronic star than the quark star.
As the rotational velocity decreases the MPR shifts outward from the centre 
of the star toward the surface.

\section{Summary and Discussion}

In this paper we have addressed the problem of path of the neutrino and the 
radius of minimum photosphere. We have done a complete GR calculation of the 
neutrino path for the most general metric describing a rotating star, and 
have obtained its geodesic equation along a given plane. 
We have calculated the MPR for four cases,
{\it i.e} stars with two different EOS and 
both rotating with two different velocities. Previous
calculation of the neutrino path was either done for a static star 
\cite{key-4,key-5}
or for a slowly rotating star \cite{key-13} and only 
along the equatorial plane. We have shown that our results also 
matches very well with
the previous findings \cite{key-4} for the Schwarzschild limit. We have found 
that the MPR is maximum along the pole and minimum along the equator. The 
MPR is much greater for the hadronic star than that of the quark star. As the
rotational velocity decreases the MPR increases and is maximum for the 
static star. Prasanna and Shrubabati \cite{key-13} had showed that the MPR 
is inversely proportional to the rotational velocity of the star. Salmonson 
and Wilson \cite{key-4} had shown that for the static star the MPR limit is
$R=3M$, and that is very close to the surface. So our results are at par 
with the previous findings in those limits and goes beyond them.

Finally we would like to mention that this calculation of neutrino path 
is very important in the sense that this forms the heart of different 
problem like GRB central engine, pulse shape and accretion disc illumination.
This path is the general path followed by any massless particle (photon) 
in the vicinity of a compact object. So the calculation might also be important 
to those problems. Currently we are trying to address other problems related
to neutrino path.

\acknowledgments{ R.M. and S.M. would like to thank CSIR for financial
support.}

\end{document}